\documentclass[reprint,amsmath,amssymb,aps,pra,]{revtex4-2}

\usepackage{graphicx}
\usepackage{dcolumn}
\usepackage{bm}
\usepackage{graphicx}
\usepackage{subfigure}
\usepackage{amsmath}
\usepackage{ulem}
\usepackage{bm}
\usepackage{amssymb}
\usepackage{color}
\usepackage{epstopdf}
\usepackage{epsfig,dsfont,multirow}
\usepackage{ulem}
\usepackage{siunitx, xcolor}
\usepackage[version=4]{mhchem}
\usepackage[colorlinks,linkcolor=blue,anchorcolor=blue,citecolor=blue]{hyperref}

\usepackage{hyperref} 
\hypersetup{
hidelinks,
colorlinks=true,
linkcolor=blue,
citecolor=blue,
anchorcolor=blue,
urlcolor = black
}

\begin{document}

\preprint{APS/123-QED}

\title{$\mathcal{PT}$-symmetry breaking and universal spectral statistics in quantum kicked rotors}
\author{Guang Li}
\affiliation{School of Science, Beijing University of Posts and Telecommunications, Beijing, 100876, CHINA}

\author{Fuxing Chen}
\affiliation{School of Science, Beijing University of Posts and Telecommunications, Beijing, 100876, CHINA}

\author{Ping Fang }
\email{pingfang@bupt.edu.cn}
\affiliation{School of Science, Beijing University of Posts and Telecommunications, Beijing, 100876, CHINA}
\affiliation{State Key Laboratory of Information Photonics and Optical Communications, , Beijing University of Posts and Telecommunications, Beijing 100876, CHINA}

\date{\today}

\begin{abstract}
We investigate the spontaneous parity-time ($\mathcal{PT}$) symmetry breaking and spectral properties of a $\mathcal{PT}$-symmetric quantum kicked rotor (QKR) under resonance conditions. At resonance, the QKR reduces to a finite-dimensional system. In the localized regime, we find that increasing the non-Hermitian parameter always induces a transition from a phase where the states exhibit $\mathcal{PT}$ symmetry to one where $\mathcal{PT}$ symmetry is spontaneously broken. In contrast, in the delocalized regime, the existence of such a transition depends on whether the reduced system is $\mathcal{PT}$-symmetric. If the reduced system is not $\mathcal{PT}$-symmetric, $\mathcal{PT}$ symmetry remains in the broken phase regardless of the non-Hermitian parameter. We further analyze the spectral statistics of the system in the delocalized regime. For real energy spectra, the level spacing distribution transitions from Wigner-Dyson statistics, associated with the Gaussian Orthogonal Ensemble (GOE), to Poisson statistics as the non-Hermitian parameter increases, with the intermediate regime well described by the Brody distribution. For complex spectra, the level spacing ratios and distributions are governed by time-reversal symmetry. The spectral statistics align with predictions for non-Hermitian random matrix ensembles in classes $\rm {AI}^\dagger$ and $\rm {A}$, depending on the presence or absence of time-reversal symmetry. Our results provide new insights into the spectral characteristics of non-Hermitian quantum chaotic systems and their connection to $\mathcal{PT}$ symmetry.

\end{abstract}

\maketitle

\hypersetup{urlcolor = blue}
\section{Introduction}
{Over the past two decades, non-Hermitian physics has garnered significant attention, particularly focusing on parity-time ($\mathcal{PT}$) symmetric quantum systems. The first analysis of a non-Hermitian quantum system with $\mathcal{PT}$ symmetry was conducted by Bender and colleagues. \cite{benderRealSpectraNonHermitian1998c,benderMakingSenseNonHermitian2007a}, has undergone extensive development across various physics domains, including optics \cite{makrisBeamDynamicsSymmetric2008,longhiBlochOscillationsComplex2009, ruterObservationParityTime2010,regensburgerParityTimeSynthetic2012}, ultracold atoms \cite{ruterObservationParityTime2010,kreibichRealizingPTSymmetric2014,graefeStationaryStatesPT2012}, and nonlinear physics \cite{konotopNonlinearWavesPT2016,miriNonlinearityinducedPTsymmetryMaterial2016}. A Hamiltonian $\hat{H}$
is  $\mathcal{PT}$-symmetric if it commutes with the combined $\mathcal{PT}$ operator, i.e., [$\hat{H}$, $\mathcal{PT}$]=0, where  $\mathcal{P}$ and $\mathcal{T}$ represent the parity and time-reversal operators, respectively. $\mathcal{PT}$ symmetry remains unbroken when all eigenstates of $\hat{H}$ are also eigenstates of the $\mathcal{PT}$ operator, leading to real eigenvalues despite the non-Hermitian nature of $\hat{H}$. In this scenario, the Hamiltonian is effectively considered quasi-Hamiltonian.\cite{PhysRevA.108.012223}. Conversely, $\mathcal{PT}$ symmetry is considered broken if there exist eigenstates of $\hat{H}$ that are not eigenstates of the $\mathcal{PT}$ operator, resulting in pairs of eigenvalues becoming complex conjugates of each other quasi-energy(QE) \cite{benderRealSpectraNonHermitian1998c,benderMakingSenseNonHermitian2007a}. This exploration into $\mathcal{PT}$-symmetric systems reveals the nuanced interplay between symmetry and spectral properties, offering insights into the fundamental aspects of quantum physics across multiple experimental platforms.
}

{The Quantum Kicked Rotor (QKR) model is a cornerstone in the field of quantum chaos, renowned for its universal and intricate dynamical behaviors and statistical properties, which have garnered extensive attention. Recent years have witnessed a surge of interest in the $\mathcal{PT}$-symmetric extension of the renowned  QKR model represents a significant advancement in the study of non-Hermitian quantum phenomena.  This extension demonstrated that chaos (i.e., the absence of dynamical localization) facilitates the emergence of the exact $\mathcal{PT}$  phase \cite{westSymmetricWaveChaos2010b}. Subsequent research on a different $\mathcal{PT}$-extended QKR model \cite{longhiLocalizationQuantumResonances2017a}, where a particle is periodically kicked by a complex crystal \cite{BENDER1999272}, revealed that  dynamical localization assists the unbroken $\mathcal{PT}$ phase. In the delocalized (quantum resonance) regime, $\mathcal{PT}$ symmetry is consistently in the broken phase,  and ratchet acceleration emerges as a hallmark of unidirectional non-Hermitian transport. Recently, considerable interest has focused on the behavioral aspects of this model’s dynamics, such as directed momentum current  \cite{zhaoDirectedMomentumCurrent2019,PhysRevResearch.6.033249,Zhao_2020}, and quantization of out-of-time-ordered correlators \cite{zhaoQuantizationOutoftimeorderedCorrelators2022a}. }

{Spectral statistics represents a significant domain of inquiry within quantum chaos. In Hermitian systems, the level spacing statistics of quasi-energy (QE) in the quantum kicked rotor adhere to the Wigner-Dyson statistics typical of the Gaussian orthogonal ensemble (GOE) within delocalized regions, while conforming to a Poisson distribution in localized regions.  This supports the Bohigas-Giannoni-Schmit (BGS) conjecture \cite{bohigasCharacterizationChaoticQuantum1984}. As interest burgeons in non-Hermitian quantum systems, the inquiry naturally extends to assessing the validity of the BGS conjecture in such cases. Recent investigations into open quantum chaotic systems have sparked renewed interest \cite{saComplexSpacingRatios2020,hamazakiUniversalityClassesNonHermitian2020b,akemannUniversalSignatureIntegrability2019a}. Specifically, two recent studies on $\mathcal{PT}$-symmetric kicked tops and dissipative quantum kicked rotors 
\cite{mudute-ndumbeNonHermitianPTSymmetric2020b,jaiswalUniversalityClassesQuantum2019} have demonstrated that their level-spacing distributions align with newly established universality classes for non-Hermitian random matrices\cite{hamazakiUniversalityClassesNonHermitian2020b,jaiswalUniversalityClassesQuantum2019}. Moreover, another statistical measure, the level spacing ratio \cite{atasDistributionRatioConsecutive2013,dusaApproximationFormulaComplex2022,garcia-garciaSymmetryClassificationUniversality2022a,prasadDissipativeQuantumDynamics2022}, emerges as a significant tool for characterizing the chaotic behavior of these systems.}

{In this work, we examined the ratio of real eigenvalues of $\mathcal{PT}$-QKR across various parameter ranges and observed the statistics of the spectral level spacing  and level spacing ratio  under different symmetries at quantum resonance.  Subsequently, we analyze the behavior of the reduced system, at this point, the Bloch number is introduced.
We found  that in localized region (where the localization length $\xi_L (\xi_L \approx k^2 /2)$ is significantly smaller than the system period $M$), $\mathcal{PT}$ symmetry is spontaneously  broken when the non-Hermitian parameter exceeds a certain threshold. 
In the  delocalized region (where the localization length $\xi_L$ is much larger than the  system period $M$),
when the  Bloch number $q=0$ or $\pi/M$, the reduced system is $\mathcal{PT}$ symmetric, the $\mathcal{PT}$ symmetry is broken once  the non-Hermitian parameter exceeds a certain threshold. And when Bloch number $q\neq 0$ and $\pi/M$, the reduced system is not $\mathcal{PT}$ symmetric, the $\mathcal{PT}$ symmetry is always broken. 

{In terms of spectral statistics,  we explore the effects of time-reversal symmetry and $\mathcal{PT}$-symmetry on the statistical properties of the spectra. We find that the spectral statistics for the complex QEs are in fact independent of the $\mathcal{PT}$-symmetry and only related to the time-reversal symmetry. However,  $\mathcal{PT}$-symmetry does affect the real QEs. The distribution of level spacing transitions from Wigner-Dyson statistics (for the Gaussian Orthogonal Ensemble, GOE) to Poisson statistics as the non-Hermitian parameter increases. The intermediate distribution can be fitted by the Brody distribution\cite{brodyStatisticalMeasureRepulsion1973,batisticIntermediateLevelStatistics2013}.}

\section{$\mathcal{PT}$-symmetric quantum kicked  {rotor}}\label{sec1}
In our investigation, we explore a generalized variant of the $\mathcal{PT}$-symmetric extension of the kicked rotor \cite{longhiLocalizationQuantumResonances2017a,fangSymmetryDynamicsUniversality2015}, which is characterized by a time-periodic Hamiltonian denoted as 
\begin{equation}
\begin{aligned}
\hat{H}(t) =& \frac{\hat{p}^2}{2 I} - \gamma \hat{p}+V(\hat{x})\sum_n \delta(t-nT),\\
V(\hat{x}) =& V_0[\cos(\hat{x}) + i\lambda \sin (\hat{x})],
\end{aligned}
\end{equation}
where $\hat{x}$ and $\hat{p}$ are, respectively, the angular operator and angular momentum operator, and $I$ is the moment of inertia. The second term can be treated as a magnetic field with strength parameter $\gamma$. The kicking potential $V(\hat{x})$
acts periodically with period $T$, and the parameter $\lambda \geq 0$ quantifies the magnitude of the potential's imaginary component, controlling the degree of non-Hermiticity in the system. This $\mathcal{PT}$ symmetric sinusoidal  potential ($V(\hat{x})$) is an important example of a complex crystal \cite{PhysRevA.84.013818,PhysRevA.81.022102,makrisBeamDynamicsSymmetric2008}. In addition, when the potential function is $V(\hat{x})$, the Hamiltonian $\hat{H}$ is time-independent and its energy spectrum and corresponding Bloch eigenfunctions have been studied in several previous papers \cite{PhysRevA.84.013818,PhysRevA.81.022102,makrisBeamDynamicsSymmetric2008,BENDER1999272}: the energy spectrum is entirely real for $\lambda  \textless 1$ (unbroken $\mathcal{PT}$ phase), and complex energy spectrum arise for $\lambda  \textgreater 1$ (broken $\mathcal{PT}$ phase). Dimensionlessly, we set $\widetilde{V}_0 = \frac{V_0 T}{I}, \widetilde{\hbar} = \frac{\hbar T}{I}, \widetilde{t}=\frac{t}{T}$ and $\widetilde{\gamma} = \gamma T$. And we take $k=\frac{\widetilde{V}_0}{\widetilde{\hbar}}$. Then the system has four parameters $\widetilde{\hbar}, \widetilde{\gamma}, \lambda$ and $k$. When $\widetilde{\gamma} = 0$ and $\lambda =0$, it returns to the standard kicked rotor \cite{IZRAILEV1990299}. Thanks to the specific form of the potential given by periodic instantaneous kicks, the quantum evolution can be expressed as stroboscopic dynamics such that the wave  function $\psi$ is given by 
\begin{equation}
\psi\left(\widetilde{t} + 1 \right) = \hat{U} \psi\left(\widetilde{t}\right),
\end{equation}
with the Floquet operator
\begin{equation}
\begin{aligned}
\hat{U} =&\exp \left(\int_0^1-\frac{i \hat{H}}{\hbar} d t \right)\\
=&\exp \left(i \frac{\widetilde{\hbar}}{2} \frac{\partial^2}{\partial x^2} + \widetilde{\gamma} \frac{\partial}{\partial x} \right) \exp \{-i k[\cos (x)+i \lambda \sin (x)]\}.
\end{aligned}
\end{equation}
In momentum representation, $\psi= \sum_l \psi_l \exp (ilx)$. 
\section{QUASIENERGY SPECTRUM AND $\mathcal{PT}$-SYMMETRY BREAKING}
{ 
In this section, we discuss the perspective of eigenvalues of the Floquet matrix.

The eigenvalues $\epsilon $ are generally called quasi-energy(QE), which can be obtained by solving the eigenequation of the Floquet operator
\begin{equation}
\hat{U}|\phi\rangle=\exp (-i \epsilon )|\phi\rangle.
\end{equation}
The $\phi \rangle$ is the eigenfunctions of $\hat{U}$. In momentum representation, $\phi\rangle = \sum_l \phi_l \exp (ilx)$. Than the eigenequations can be written as 
\begin{equation}
\begin{gathered}
\sum_{n=-\infty}^{\infty} \mathcal{U}_{l, n} \phi_n (\epsilon) = \exp (-i \epsilon ) \phi_l(\epsilon),
\end{gathered}
\end{equation}
}
where $\mathcal{U}_{l, n}$ is the matrix elements of $\hat{U}$ operators in momentum representation 

\begin{equation}
\begin{aligned}
\mathcal{U}_{l, n}=&\exp \left(-i \frac{\widetilde{\hbar}}{4} l^2 + i \frac{\widetilde{\gamma}}{2}l\right)W_{l-n}\exp \left(-i \frac{\widetilde{\hbar}}{4} n^2 + i \frac{\widetilde{\gamma}}{2}n\right),
\end{aligned}
\end{equation}
and $W_n$ is the Fourier coefficient of $\exp \{-i k[\cos (x)+i \lambda \sin (x)]\}$, i.e., $\exp \{-i k[\cos (x)+i \lambda \sin (x)]\}=\sum_n W_n \exp (i n x)$. The relation between $\psi$ and $\phi$ is $\phi_l=exp(i\epsilon \widetilde{t})\psi_l$.

The current work considers the case of quantum resonance($\widetilde{\hbar}=4\pi N/M$, where $N$ and $M$ are coprime positive integers). When $\widetilde{\gamma}=0$, the Floquet operator has a period such that 
$\mathcal{U}_{n+M, l+M}=\mathcal{U}_{n, l}$. For $\widetilde{\gamma}=2 \pi a / (bM)$,  where $a$ and $bM$ are coprime positive integers, the Floquet operator satisfies
$\mathcal{U}_{n+bM, l+bM}=\mathcal{U}_{n, l}$.
{ 
Employing Bloch's theorem, the eigenfunctions can be written as
\begin{equation}
\phi_l=c_l \exp (i q l), \quad c_{l+bM}=c_l,
\end{equation}
$q \in (-\pi/(bM),\pi /(bM)]$ is Bloch number. In this case, the eigenequation for the Floquet operator can be reduced
}
\begin{equation}
\begin{gathered}
\sum_{n=0}^{b\times M-1} S_{l, n} c_n=\exp (-i \epsilon ) c_l,
\end{gathered}
\end{equation}
where the $(bM)\times (bM)$ matrix coefficients  $S_{l, n}$ are defined by
\begin{equation}
\begin{aligned}
S_{l, n}(q)=&\exp \left[- i \frac{\widetilde{\hbar}}{4} l^2+i \frac{\widetilde{\gamma}}{2}l\right]\sum_{j=-\infty}^{\infty} W_{l-j b M-n} \\
  &\times\exp \left[i q(j bM+n-l)- i \frac{\widetilde{\hbar}}{4} n^2+i \frac{\widetilde{\gamma}}{2}n\right].
\end{aligned}
\end{equation}
Hence the QE $\epsilon(q)$ can be obtained through the eigenvalues $\exp (-i \epsilon(q) )$ of the $bM\times bM$ matrix $S_{l, n}(q)$. 
{It can be demonstrated that for $q=0$ or $\pi/(bM)$,  the Floquet operator exhibits the symmetry 
 \begin{equation}
     C_{\mathcal{PT}} S C_{\mathcal{PT}}^{-1} =S^{-1}.
 \end{equation}
}
{The physical meaning of this symmetry is the conservation of ${\mathcal{PT}}$ symmetric for the system under the transformation $t \rightarrow-t$ together with $\theta \rightarrow-\theta$ \cite{IZRAILEV1990299}.}

Besides that, when $\gamma=0$ and $2 \pi L / M$  and  $L$ is  an integer coprime to $M$ the Floquet operator $S$ has the symmetry
\begin{equation}
   \mathcal{S}_{l, n}=\mathcal{S}_{-n+p, -l+p},
 \end{equation}
 $p=(L\times\Delta)\bmod M,$ where $\Delta=(M+1)/N$, and $\Delta$ and $d$ are integers. And the system in this case has time-reversal symmetry \cite{IZRAILEV1990299}. Conversely, when $b>1$ , the system lacks time-reversal symmetry.
\begin{figure}[ht]
\centering
\includegraphics[width=8.6cm]{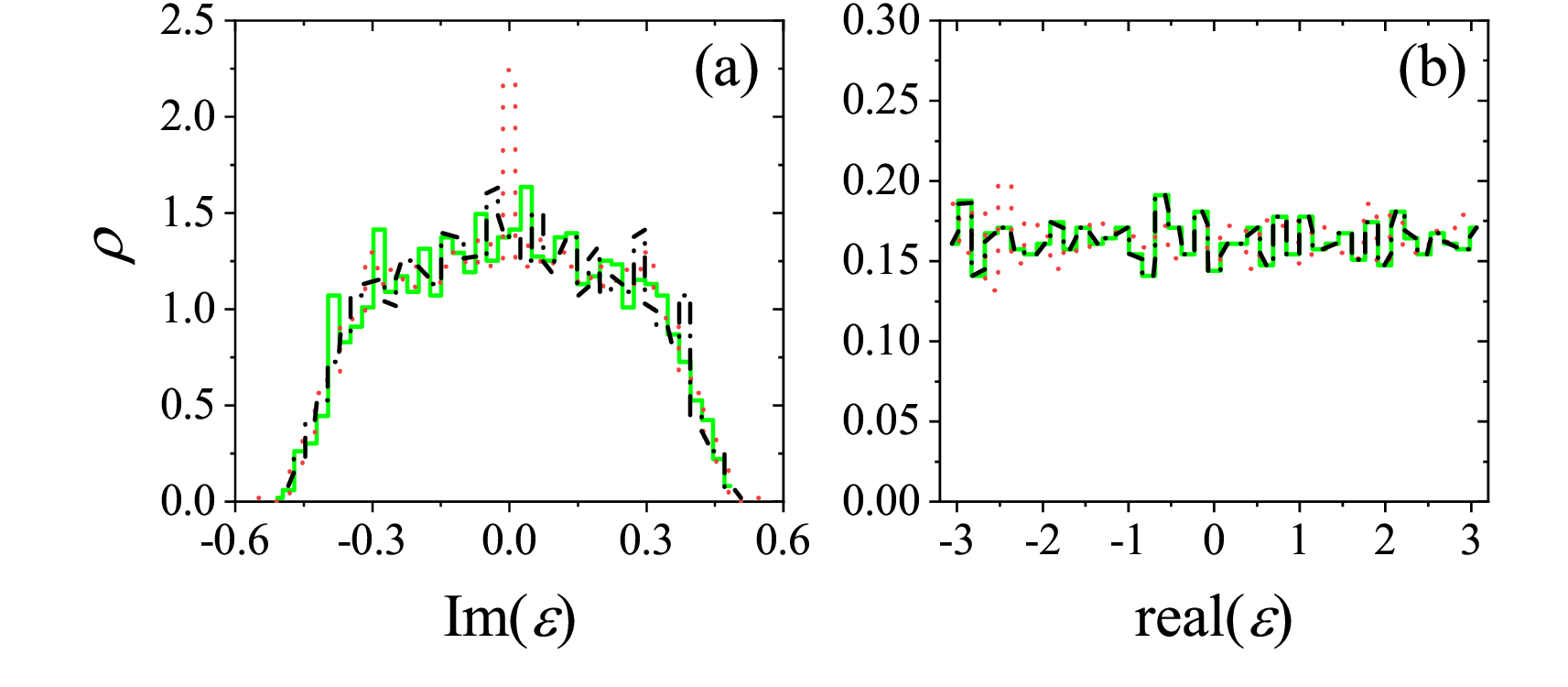}
\caption{Density distribution of the real and imaginary  parts of the QE at $k=10^5, \lambda=10^{-5}$ and $\hbar = 4\pi /399$ for $q=-0.001$ (green solid histograms), $q=0$(red dotted histograms) and $q=0.001$ (black dash-dotted histograms). }\label{fig11}
\end{figure}

\begin{figure}[ht]
  \centering
 \includegraphics[width=8.6cm]{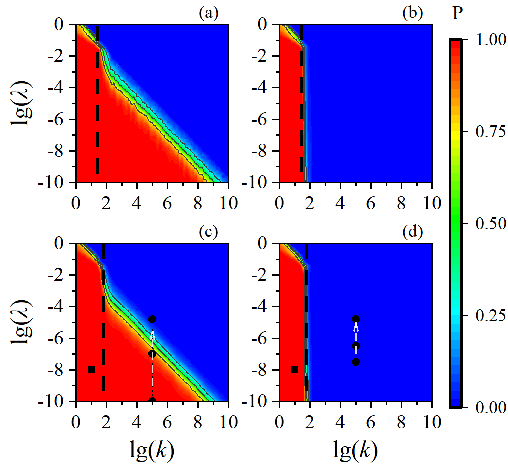}
  \caption{{The phase diagrams for $q = 0$ (a) and $q = 0.001$ (b) at $\widetilde{\hbar} = 4 \pi /399$ and $q = 0$ (c) and $q = 0.001$ (d) at $\widetilde{\hbar} = 5\times4 \pi /1999$. The black dashed line in the phase diagram is the value of $k$ when the corresponding localization length {$\xi_L (\xi_L \approx k^2 /2)$ } is equal to the period $M$ of the system.} The statistical nature of the level statistics at the black dots will be elucidated in detail in Fig.~\ref{fig3}. In addition, the parameter ranges in Fig.~\ref{fig6} (a) are at the white arrows.}\label{fig2}

\end{figure}

{ 
After obtaining the QEs numerically at $\mathcal{PT}$-symmetry broken phase, we plot the density distributions, denoted as $\rho$, of the real and imaginary  parts of the QEs [Fig.~\ref{fig11}]. Our analysis reveals that the real part of the QEs exhibits a uniform distribution, consistent with findings observed in Hermitian kicked rotors \cite{IZRAILEV1990299}. Conversely, the distribution pattern of the imaginary  part closely resembles that observed in non-Hermitian kicked tops \cite{mudute-ndumbeNonHermitianPTSymmetric2020b}. Consequently, it is necessary to unfold solely the imaginary  part when investigating the level spacing distributions. Additionally, the QEs demonstrate a complex conjugate symmetry about $q$, expressed as $(\epsilon(-q) = \epsilon^*(q))$ \cite{longhiLocalizationQuantumResonances2017a}.
}

{ 
We define the ratio of real quantum eigenvalues as $\rm P$. If $\rm P$ is less than $1$, then the $\mathcal{PT}$-QKR resides in a broken $\mathcal{PT}$-symmetry phase. We performed numerical calculations by varying parameters $k$ and $\lambda$, and the phase diagrams at $\gamma=0$ are presented in Fig.~\ref{fig2}. Although our research is conducted under resonance conditions, when $ k^2 /2 \ll M$,  the system  enters an effective localized state, and the localization  length $\xi_L  \approx k^2 /2$ \cite{IZRAILEV1990299}.  This condition remains valid even in non-Hermitian systems \cite{westSymmetricWaveChaos2010b}.  In the localized regime,  localization (where the localization length $\xi_L$ is much smaller than the system period $M$) preserves $\mathcal{PT}$ symmetry, although regions with large values of $k\times\lambda$ still exhibit $\mathcal{PT}$ symmetry breaking. This is similar to the role of dynamical localization in promoting $\mathcal{PT}$ symmetry under non-resonant conditions($\widetilde{\hbar} \neq 4\pi N/M$) \cite{longhiLocalizationQuantumResonances2017a,Sharma:2024fqc}. In the delocalized regime,  the presence or absence of $q = 0$ or $\pi/M$ significantly influences $\mathcal{PT}$ symmetry breaking. $\mathcal{PT}$ symmetry breaking occurs in regions where $k\times\lambda$ is large when $q = 0$ or $\pi/M$. Conversely, for $q \neq 0  $ and $q\neq\pi/M$, the system consistently exhibits $\mathcal{PT}$ symmetry breaking. When $q = 0$ or $\pi/M$, the reduced system is $\mathcal{PT}$ symmetric. Therefore, in the case of delocalized regime, the $\mathcal{PT}$ symmetric of the reduced system will affect the properties of $\mathcal{PT}$ symmetry breaking.

\begin{figure}[ht]
  \centering
  \includegraphics[width=8.6cm]{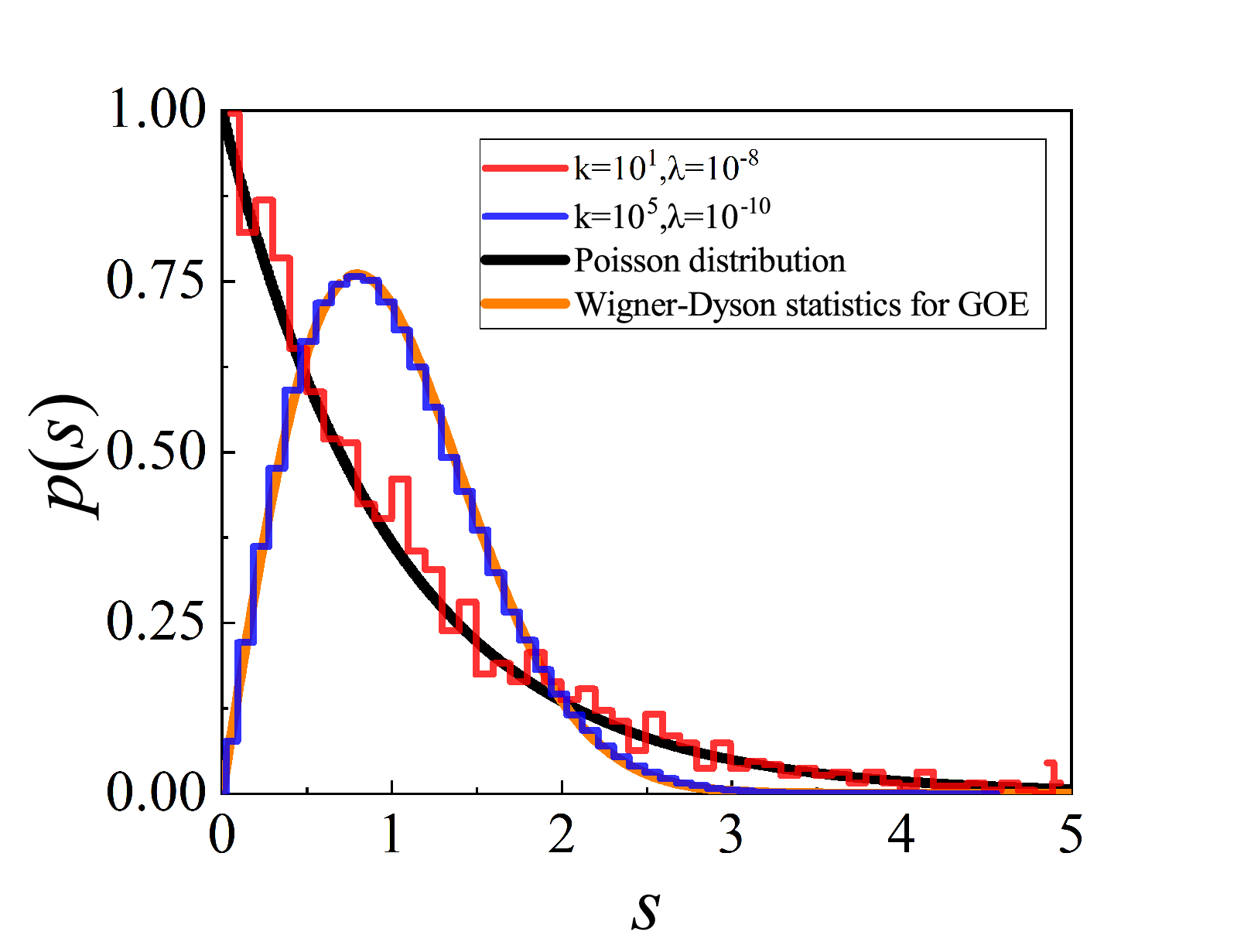}
  \caption{(Color online) Level-spacing distributions of the systems in the quasi-Hermitian case. In the localized region the distribution (red histograms) follow the Poisson distributions (black line)  by averaging 200 Bloch values taken uniformly from $[0, \pi / M)$, $k=10^{1}, \lambda =10^{-8}$ and $\widetilde{\hbar} =5 \times 4 \pi /1999$.
  And in the delocalized region the distribution (blue histograms) follow  the Wigner-Dyson statistics for GOE (orange line) by averaging $k\in[10^{5}-50, 10^{5}+50]$, with the
interval $\Delta k=1$ ,$\lambda=10^{-10}$, and $\widetilde{\hbar} =5\times 4 \pi /1999$}.
\label{fig9}
\end{figure}
 \section{Level statistics}\label{sec2}

In this section, we compute the distributions of level spacing and level-spacing ratios. In the Hermitian case, the level spacing statistics of the QEs of the kicked rotor follow the   Wigner-Dyson statistics for GOE in the delocalized region and the Poisson distribution in the localized region, consistent with the Bohigas-Giannoni-Schmit (BGS) conjecture. However, in the non-Hermitian case, the QEs become complex. We analyze the distributions of Euclidean nearest neighbor distances in the complex plane. Before calculating spectral statistics, unfolding is required. Since the real part of the QEs is uniformly distributed, unfolding is only necessary for the imaginary  part. We utilize the integrated staircase function of the imaginary  parts to approximate the underlying smooth distribution for unfolding \cite{abul-magdUnfoldingSpectrumChaotic2014}. It is worth noting that a statistical approach without unfolding, namely level-spacing-ratio distributions, has been recently proposed \cite{oganesyanLocalizationInteractingFermions2007,dusaApproximationFormulaComplex2022,garcia-garciaSymmetryClassificationUniversality2022a,prasadDissipativeQuantumDynamics2022}.
{The level-spacing-ratio 
\begin{equation}
z_m=\frac{\epsilon_m^{\mathrm{NN}}-\epsilon_m}{\epsilon_m^{\mathrm{NNN}}-\epsilon_m},
\end{equation}
where $\epsilon _m^{NN}$ and $\epsilon _m^{NNN}$ are the nearest and next-to-nearest neighbor of $\epsilon _m$ in the complex plane. Since the QEs is isotropic, we take $r=\mid z_m\mid$.}

{We counted the level-spacing distributions of the real QEs and complex QEs and the real part of QEs, respectively. The level-spacing $ \{s\}$ is the Euclidean distance between two nearest neighboring points on the complex plane or real axis.}

{The statistic of real QEs in the $\mathcal{PT}$-symmetry phase is shown in Fig.~\ref{fig9}.} In the localized region ($\xi_{L}<M$), when the $\mathcal{PT}$-QKR is in the $\mathcal{PT}$-symmetry unbroken phase ({indicated by} black square point in Fig.~\ref{fig2} (c) and (d).), the level-spacing distributions of the system follow a Poisson distribution, and in the {delocalized} region ($\xi_{L}>M$), when the $\mathcal{PT}$-QKR is in the $\mathcal{PT}$-symmetry unbroken phase (indicated by bottom black dot in Fig.~\ref{fig2} (c).), the level-spacing distributions of the system follow the Wigner-Dyson statistics for GOE.

\begin{figure}[ht]
\centering
\includegraphics[width=8.6cm]{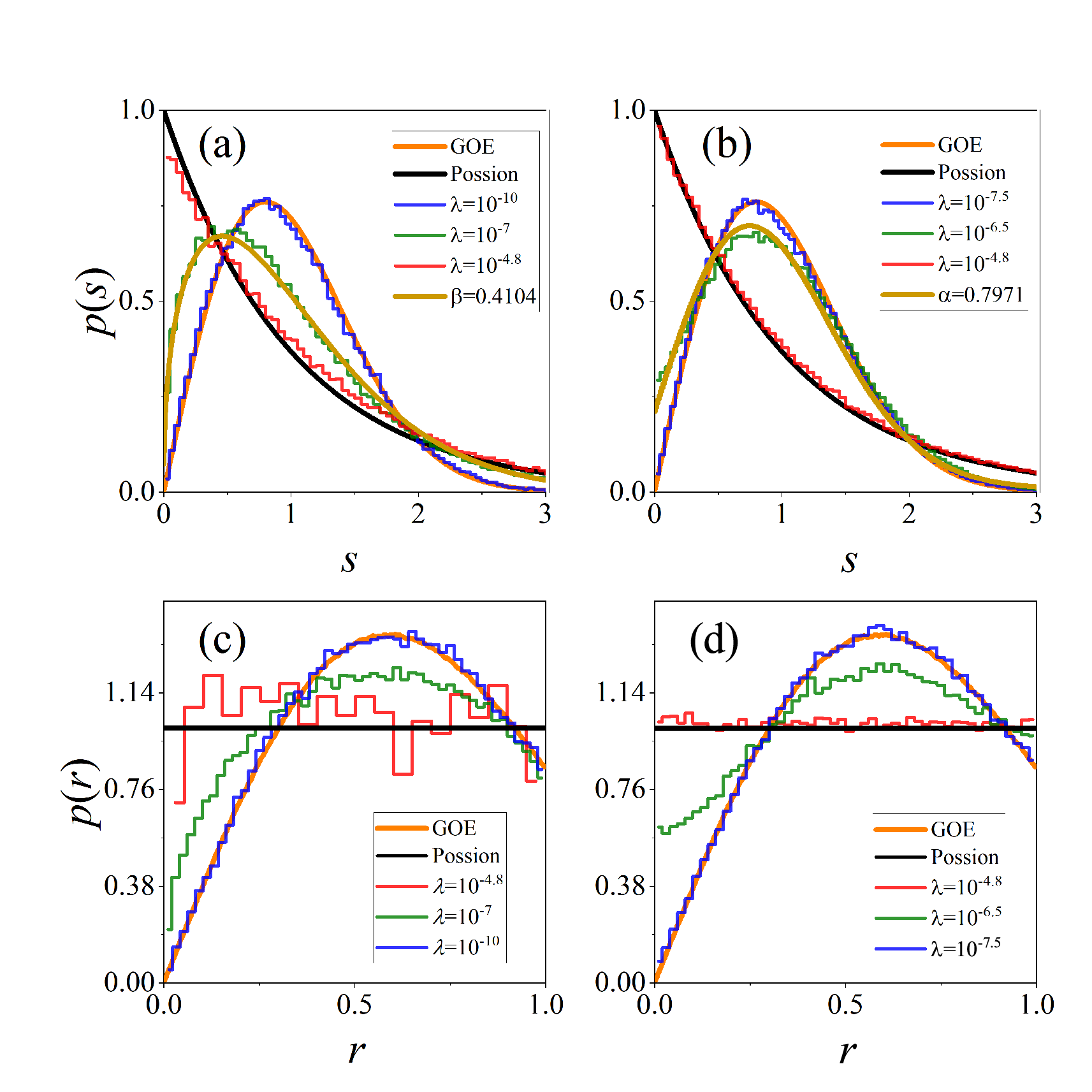}
\caption{(Color online) Level-spacing distributions of the real QEs at $q = 0 (a)$ and the real part of the QEs at $q=0.001 (b)$ and the distributions of level spacing ratio $r$ for of the real QEs at $q = 0 (c)$ and the real part of the QEs at $q=0.001 (d)$, in the delocalized region by averaging $k\in [10^{5}-50, 10^{5}+50]$, with the interval $\Delta k = 1$ and $\widetilde{\hbar} =5 \times 4 \pi /1999$.  As $\lambda$ increases (the direction of the white arrow in Fig.~\ref{fig2} (c) (d)), the distribution of both transition from the Wigner-Dyson statistics for GOE (orange line) to Poisson (black line)). 
{And the distribution are shown in $(a)$ and $(c)$ for $\lambda=10^{-10}$ (blue  histograms), $\lambda=10^{-7}$ (green  histograms) and $\lambda=10^{-4.8}$ (red line), and in $(b)$ and $(d)$ for $\lambda=10^{-7.5}$ (blue histograms), $\lambda=10^{-6.5}$ (green  histograms) and $\lambda=10^{-4.8}$ (redline), respectively.}}\label{fig3}
\end{figure}

Then, we investigate the transitional behavior of the level spacing distribution of the real QEs in the delocalization region (for $q = 0$ or $\pi/M$). Conversely, for $q \neq 0$ and $q\neq\pi/M$, considering that the QEs are complex numbers at this point, we analyze the real part of the QEs. As depicted in Fig.\ref{fig3} (a), when $q = 0$ or $\pi/M$, the level spacing distribution transitions from the Wigner-Dyson statistics for GOE to Poisson distribution as $\lambda$ increases (as indicated by the white arrow in Fig.~\ref{fig2} (c)). The intermediate states can be well fitted by the Brody distribution\cite{brodyStatisticalMeasureRepulsion1973,batisticIntermediateLevelStatistics2013}, represented as 
\begin{equation}
P_B(s)=C_1 s^\beta \exp \left(-C_2 s^{\beta+1}\right),
\end{equation}
where $C_1$ and $C_2$ are determined by $\langle s\rangle=1$,
\begin{equation}
C_1=(\beta+1) C_2, \quad C_2=\left(\Gamma\left(\frac{\beta+2}{\beta+1}\right)\right)^{\beta+1}
\end{equation}
with $\Gamma(x)$ being the Gamma function.The fitting parameter $\beta$ is in the interval $[0,1]$, where $\beta=0$ corresponds to Poisson distribution, and $\beta=1$ gives the Wigner surmise. And $\beta=0.4104$ when $\lambda=10^{-7}, k=10^{5}$. Furthermore, the results in Fig.\ref{fig1} indicate that the fitting parameter $\beta$ is only related to the value of $k\times\lambda$. For $q \neq 0$ and $q\neq\pi/M$, when the non-Hermitian parameter $\lambda$ is very small, the level spacing distribution of the real parts of the QEs follows the Wigner-Dyson statistics for GOE. As $\lambda$ increases (as shown by the white arrow in Fig.~\ref{fig2} (d)), the distribution transitions from the Wigner-Dyson statistics for GOE to Poisson, and the intermediate states can be fitted by a linear combination of Poisson and the Wigner-Dyson statistics for GOE $P(s)=\alpha\times P_{GOE}(s)+(1-\alpha)\times P_{Poisson}(s)$, where $P_{GOE}(s)$ and $P_{Poisson}(s)$ refer to Wigner-Dyson statistics for GOE and Poisson distribution, respectively. As shown in Fig.~\ref{fig3}(b), when $\lambda = 10^{-6.5}$, $\alpha$ takes $0.7971$.

\begin{figure}[ht]
\centering
\includegraphics[width=8.6cm]{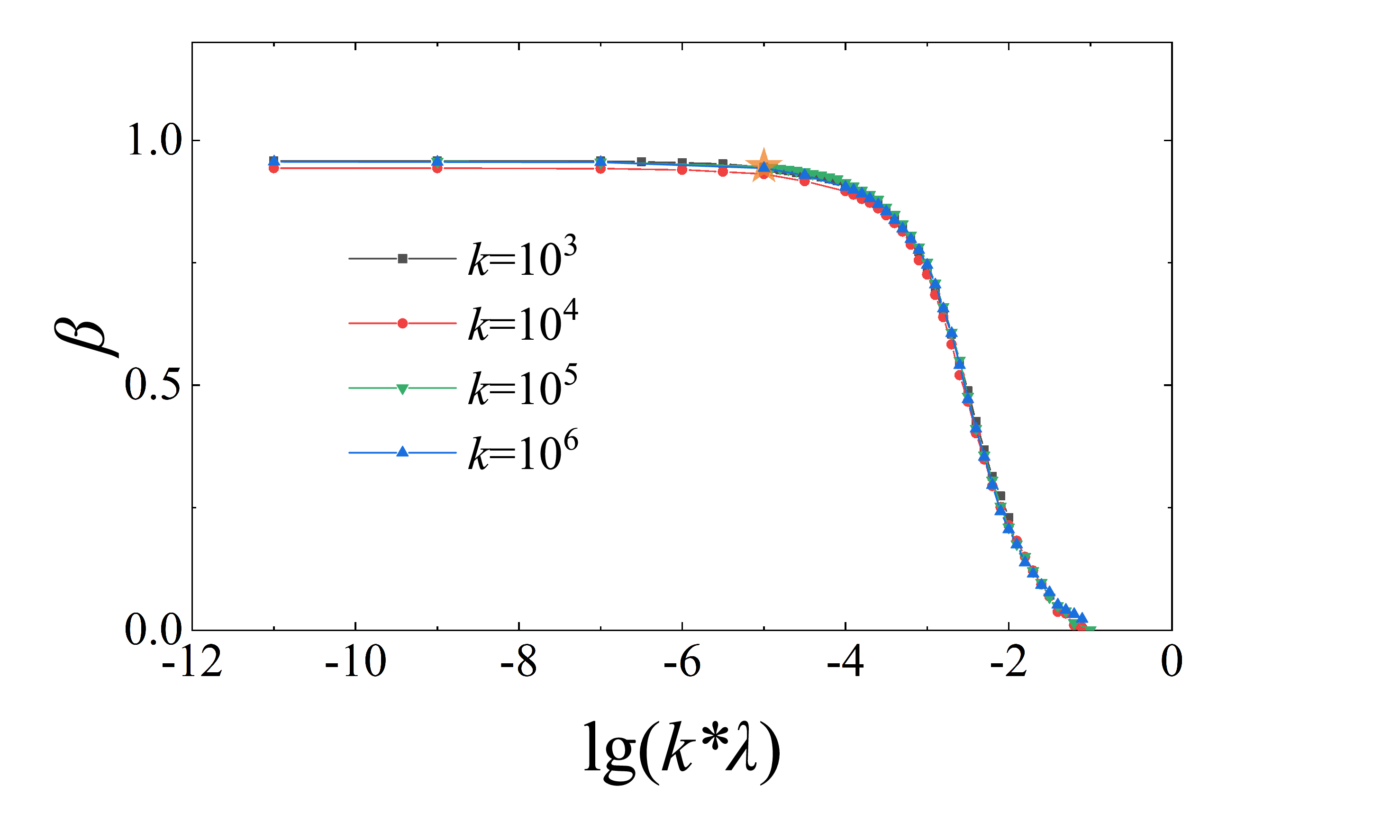}

\caption{Results for the Brody parameter $\beta$ with $k\times\lambda$ at different $k$ and $\hbar = 5\times4\pi /1999$. The yellow pentagram indicates the distribution of red dash-dotted histograms in Fig.~\ref{fig9}.}\label{fig1}
\end{figure}

\begin{figure}[ht]
\centering
\includegraphics[width=8.6cm]{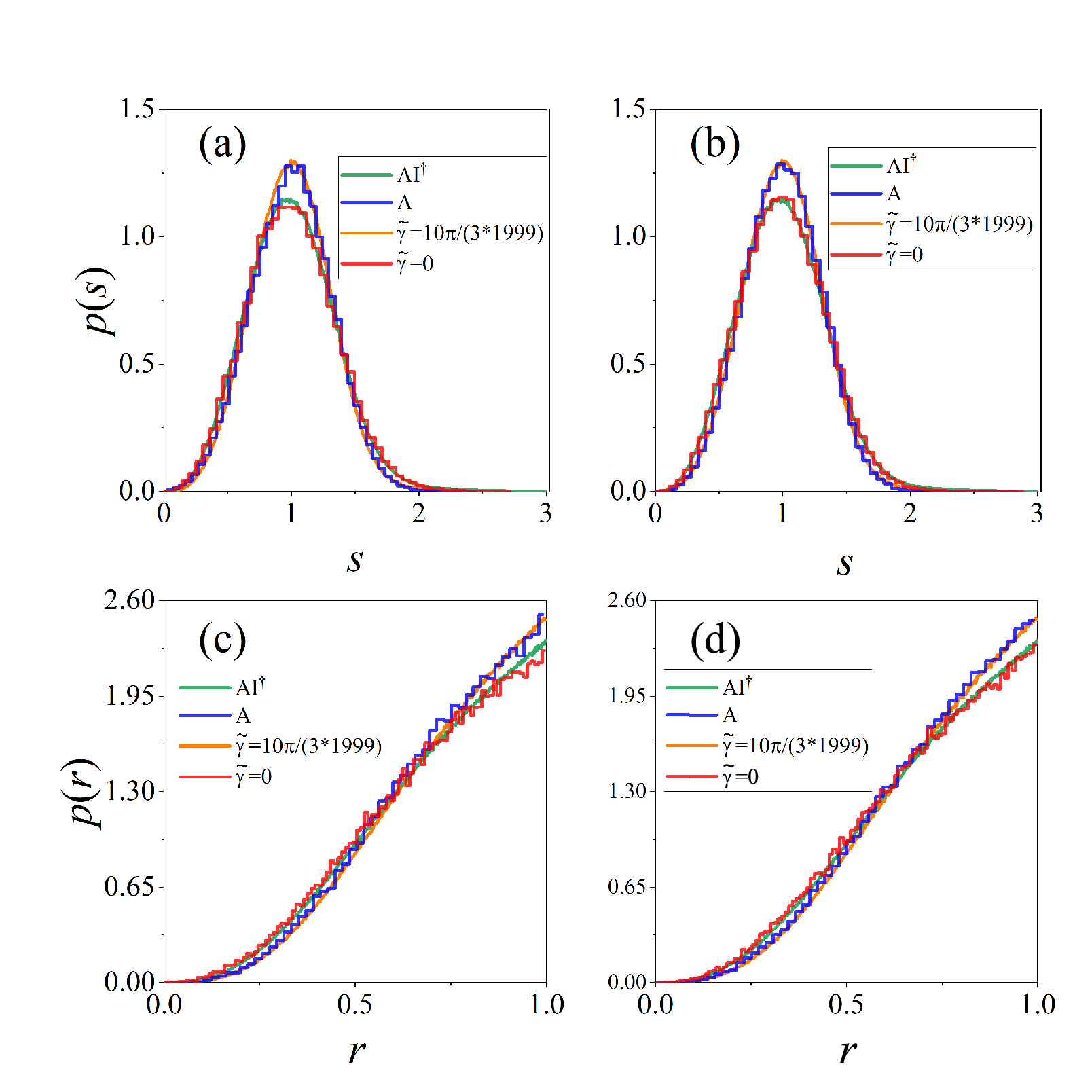}

\caption{{(Color online) The level spacing distributions of complex QE in the delocalized region at $q=0$ (a), $q=0.001(b)$ and  the distributions of level spacing ratio $r$ of complex QE  in the delocalized region at $q=0 (c)$, $q=0.001(d)$ by averaging $k\in[10^{5}-50, 10^{5}+50]$, with the interval $\Delta k=1$, $\widetilde{\hbar} =5\times 4 \pi /1999$. When $\widetilde{\gamma}=10\pi/(3\times1999)$, the simulation results of the distribution (blue histograms) correspond to the universality classes of the random-matrix ensembles in class $\rm {A}$ 
 (orange line). While $\widetilde{\gamma}=0$, the simulation results of the distribution (red  histograms) correspond to the universality classes of the random-matrix ensembles in class $\rm {AI}^\dagger$  
(green line). These results from non-Hermitian random matrix theory are derived from the diagonalization of $5,000$ matrices, each with dimensions of 3000 × 3000.}}\label{fig4}
\end{figure}

\begin{figure}[ht]
 \centering
 \includegraphics[width=7.5cm]{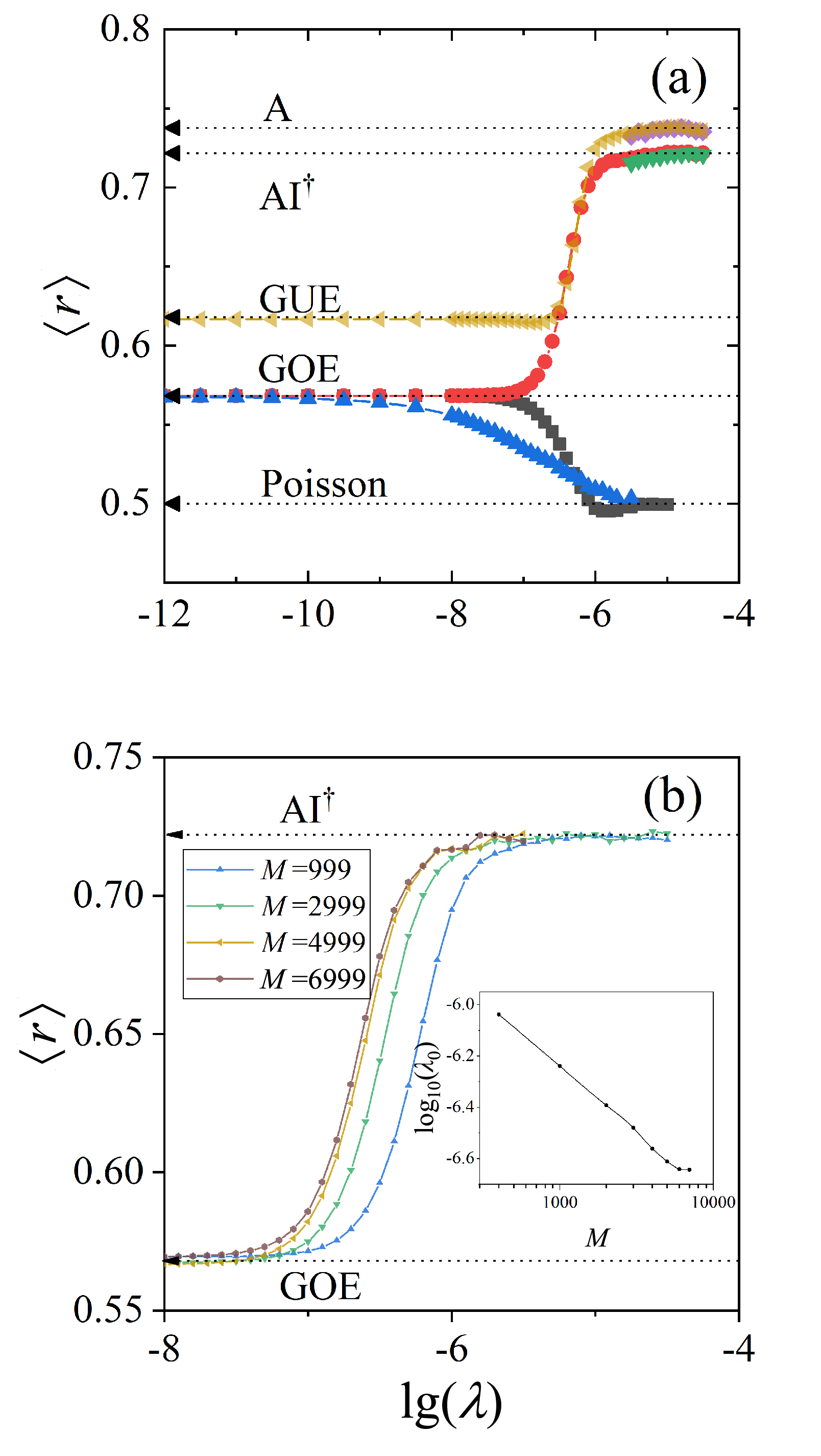}
 \caption{{The $\langle r\rangle$ varies with $\lambda$ in the delocalized region. (a) complex QEs with different parameters [ $q=0, \gamma=0$ (green inferior triangle); $q=0.001, \gamma=0$ (red solid line with circle); $q=0, \gamma=$ $10 \pi /(3 \times 1999)$ (purple rhombic) and $q=0.001, \gamma=10 \pi /(3 \times$ 1999) (orange solid line with left triangle)], real QEs at $q=0, \gamma=0$ (blue solid line with upper triangle), and real part of the QEs at $q=$ $0.001, \gamma=0$ (black square) by averaging $k \in\left[10^5-50,10^5+50\right]$, with the interval $\Delta k=1$, and $\hbar=5 \times 4 \pi / 1999$. (b) the effect of matrix size (M) on the transition of $\langle r\rangle$ for the case of the red solid line with circle in (a).}}\label{fig6}
\end{figure}

In non-Hermitian random matrix theory, research has identified only three nearest-neighbor spacing distributions for random matrices that exhibit $\rm {A}$, $\rm {AI}^\dagger$, and $\rm {AII}^\dagger$ symmetries among the nine distinct single symmetries \cite{hamazakiUniversalityClassesNonHermitian2020b}. Non-Hermitian random matrices in class $\rm {A}$ have no symmetry constraint; matrices in class $\rm {AI}^\dagger$ respect $H=H^T\left(\neq H^*\right)$, Regarding the spacing distribution, recent work has proposed some theoretical approximations \cite{PhysRevE.106.014146}.

{More notably, the statistics of the complex QEs in the phase of $\mathcal{PT}$ symmetry breaking for $\mathcal{PT}$-QKR depend solely on the system's time-reversal symmetry. Through numerical analysis, we examine the level spacing distributions for $\mathcal{PT}$-QKR with differing symmetries and juxtapose these with outcomes derived from simulations of non-Hermitian random matrices. As shown in Fig.\ref{fig4} , for cases where $\widetilde{\gamma} =10\pi/(3\times1999)$,  which indicates the absence of time-reversal symmetry, the simulation outcomes (blue histograms) correspond to the universality classes of the random-matrix ensembles in class $\rm {A}$  (orange line). Conversely, when $\widetilde{\gamma} = 0$, signifying the presence of time-reversal symmetry, the simulated distributions (red histograms) correspond to the universality classes of the random-matrix ensembles in class $\rm {AI}^\dagger$
(green line).
}}

{Finally, we plot the mean value of $r$ as a function of $\lambda$ in Fig.~\ref{fig6} (a) in the delocalized regime. The results show that the values of $\langle r\rangle$ of QEs is related only to system symmetry.  When $q\neq0$ and $q\neq\pi/bM$, as $\lambda$ increases, at $\gamma=0$ (with time-reversal symmetry), the values of $\langle r\rangle$ of  complex QEs transition from Wigner-Dyson statistics for GOE ($\langle r\rangle=0.5687$) to the universality classes of the random-matrix ensembles in class $\rm {AI}^\dagger$ ($\langle r\rangle=0.7218$); at $\gamma=10\pi/(3\times1999)$ (without time-reversal symmetry), the values of $\langle r\rangle$ of QEs transition from Wigner-Dyson statistics for the Gaussian Unitary Ensemble (GUE) ($\langle r\rangle=0.6180$) to the universality classes of the random-matrix ensembles in class $\rm {A}$ ($\langle r\rangle=0.7378$).  These results from non-Hermitian random matrix theory are derived from the diagonalization of $5,000$ matrices, each with dimensions of 3000 × 3000.  And when $q=0$ or $\pi/M$, as $\lambda$ increases the values of $\langle r\rangle$ of real QEs transition from Wigner-Dyson statistics for GOE to Poisson at $\gamma=0$, 
and the values of $\langle r\rangle$ of complex QEs is correspond to the universality classes of the random-matrix ensembles in class $\rm {A}$ at $\gamma=10\pi/(3\times1999)$ and in class $\rm {AI}^\dagger$ at $\gamma=0$ at larger $\lambda$.} These transitions indicate that the system shifts from quasi-Hermitian to non-Hermitian. However, these transitions do not constitute phase transitions. We find that as the matrix size increases, the change in $\langle r\rangle$ tends to stabilize  rather than become sharp. The results of the variation of  $\langle r\rangle$  at $\gamma = 0, q =0.001$ for different matrix sizes are shown in Fig.~\ref{fig6} (b). Additionally,  $\lambda_0$ is the value of $\lambda$ corresponding to the average of the values of $\langle r\rangle$ of the universality classes of the random-matrix ensembles in class $\rm {AI}^\dagger$ and Wigner-Dyson statistics for GOE, which is 0.6425. As $M$ increases, $\lambda_0$ gradually converges to a constant value.

\section{conclusion}

{In summary, our investigation focused on the spectral characteristics of $\mathcal{PT}$-symmetric Quantum Kicked Rotors ($\mathcal{PT}$-QKR) under various parameters, with a particular emphasis on spectral statistics (level spacing and level spacing ratio statistics) across different symmetries. We discovered that in the delocalized region, where the localization length $\xi_L$ significantly exceeds the system's period (M), $\mathcal{PT}$-symmetry breaks as $k\times\lambda $ increases for Bloch numbers 
$q=0$ or $\pi/M$; whereas for $q\neq0$ and $q\neq\pi/M$, the $\mathcal{PT}$-QKR consistently remains in a phase of broken $\mathcal{PT}$-symmetry. Conversely, in the localized region, where $\xi_L$ is much smaller than M, the $\mathcal{PT}$-QKR stays in the unbroken $\mathcal{PT}$-symmetry phase, despite the possibility of $\mathcal{PT}$-symmetry breaking in areas with high  $k\times\lambda $ values. This finding aligns with previous conclusions, indicating that dynamical localization supports the maintenance of $\mathcal{PT}$-symmetry in resonant scenarios.
}

{
Regarding spectral statistics, within the localized region and when the system is in the $\mathcal{PT}$ symmetry phase, the level spacing distribution follows a Poisson distribution. In the delocalized region, for $q=0$ or 
$\pi/M$, as $k\times\lambda $ increases, the level spacing distribution for real QEs transitions from Wigner-Dyson (GOE) statistics to a Poisson distribution, with intermediate states well approximated by the Brody distribution \cite{brodyStatisticalMeasureRepulsion1973,batisticIntermediateLevelStatistics2013}. For 
$q\neq0$ and 
$q\neq\pi/M$, the level spacing distribution for the real part of QEs moves from Wigner-Dyson (GOE) statistics to a Poisson distribution, with intermediate states represented by a mixture of Poisson and Wigner-Dyson (GOE) statistics. Furthermore, the statistics of level spacing ratios and the distribution of spacings in the complex energy spectrum solely depend on the presence of time-reversal symmetry. As the non-Hermitian strength $\lambda$ increases (with $k$ constant), the system transitions from Wigner-Dyson (GOE) statistics  to the universality classes of the random-matrix ensembles
in class $\rm {AI}^\dagger$   in the presence of time-reversal symmetry, and from Wigner-Dyson (GUE) statistics  to the universality classes of the random-matrix ensembles
in class $\rm {A}$  in the absence of time-reversal symmetry.
}

{Our results further show the connection between  non-Hermitian random matrix theory and quantum chaos. In addition, for real spectra, we  have the BGS conjecture that associates dynamics with spectral correlations. However, it remains unclear to what extent the consistency with non-Hermitian random matrix predictions is related to quantum chaos in the sense of quantum dynamics of classically chaotic systems. In addition, whether the system has time-reversal symmetry at $\gamma= 
2 \pi L / M$ in the resonance case has to be further verified experimentally.}

\section{Acknowledgements}
We would like to thank Prof. Zheng Zhu  for the discussions.
This research was supported by the Fundamental Research Funds for the Central Universities with contract number 2024ZCJH13,the National Natural Science Foundation
of China (Grant No. 12005024), the Fundamental Research Funds for the Central Universities(No. 2023ZCJH02),
and the High-performance Computing Platform of Beijing
University of Posts and Telecommunications.

\end{document}